\documentclass[prd,twocolumn]{revtex4}
\usepackage{dcolumn}
\usepackage{multirow}
\usepackage{graphicx}
\usepackage{amssymb}
\usepackage{bm}
\usepackage{hyperref}
\usepackage{epstopdf}
\usepackage{color}
\usepackage{mathrsfs}
\usepackage{amsmath,amssymb,amsthm}
\usepackage{rotating}
\usepackage{sverb, longtable}
\usepackage{subfigure}

\usepackage[graphicx]{realboxes}
\usepackage{adjustbox}
\begin{document}

\title{Quintessence Dark Matter}
\author{Deng Wang}
\email{dengwang@ific.uv.es}
\affiliation{Instituto de F\'{i}sica Corpuscular (CSIC-Universitat de Val\`{e}ncia), E-46980 Paterna, Spain}

\begin{abstract}
Recently, we give the robust $\sim2\,\sigma$ evidences of dynamical dark matter and beyond $2\,\sigma$ signals of the coexistence of dynamical dark matter and dynamical dark energy using current cosmological observations \cite{Wang:2025zri}. Here we propose the quintessence dark matter model to explain the evolution of dark matter over time on cosmic scales. Interestingly, we find that the exponential quintessence is likely the origin of such an evolution.

\end{abstract}
\maketitle

\section*{\normalsize I. Introduction}
Over a considerable span, the nature of dark energy (DE) is believed to be the cosmological constant, since the cosmic acceleration is discovered by two type Ia supernova teams \cite{SupernovaSearchTeam:1998fmf,SupernovaCosmologyProject:1998vns}. In 2024, the DESI collaboration \cite{DESI:2024mwx} reported a strong evidence of dynamical dark energy (DDE) based on the combination of cosmic microwave background (CMB) \cite{Planck:2018vyg,ACT:2023kun}, type Ia supernova (SN) \cite{Brout:2022vxf,Rubin:2023ovl,DES:2024jxu} and their high-precision measurements of baryon acoustic oscillations (BAO) \cite{DESI:2024uvr,DESI:2024lzq}. Recently, using the same analysis methods, this DDE evidence is further enhanced in light of their data release two (DR2) \cite{DESI:2025zgx,DESI:2025fii,DESI:2025zpo}. However, we question the validity of their DDE analysis due to the fact that the data combination is problematic \cite{Wang:2025bkk}. The strong evidence of the evolution of DE suffers from the tensions among CMB, BAO and SN datasets. More interestingly, using eight individual datasets including CMB, DESI DR2 BAO and SN observations, we demonstrate that today's universe could be in a decelerating phase in the statistically preferred Chevallier-Polarski-Linder (CPL) \cite{Chevallier:2000qy,Linder:2002et} DDE scenario over the $\Lambda$CDM model \cite{Wang:2025owe}. These new findings will lead to a reevaluation of DE and cosmic evolution. 

It is worth noting that all the findings including cosmic acceleration and DDE are obtained by assuming a completely cold dark matter (CDM) in the cosmic pie, i.e., the equation of state (EoS) of dark matter (DM) $\omega_{dm}=0$. It is natural to ask if DM like DE also evolves over time. Most recently, we find the robust $\sim2\,\sigma$ evidences of dynamical dark matter (DDM) and beyond $2\,\sigma$ signals of the coexistence of DDM and DDE based on current observations \cite{Wang:2025zri}. But, we do not explain the possible nature of DDM in theory. 

In \cite{Wang:2025zri}, we aim at implementing the data analyses to find DDM evidences. In this study, we focus on explaining the properties of DDM more physically. Specifically, we propose the quintessence dark matter (QDM) model to interpret the evolution of DM on cosmological scales. We point out that the redshift evolution of DM is likely originated from the exponential quintessence.

\section*{\normalsize II. QDM}
In the 1980s, the cosmological dynamics of quintessence in the presence of matter and radiation have been extensively studied in \cite{Fujii:1982ms,Ford:1987de,Wetterich:1987fm,Ratra:1987rm}. Until 1997, the term ``quintessence'' is firstly coined by Caldwell {\it et al.} \cite{Caldwell:1997ii} to represent a canonical scalar field $\tilde{\phi}$ with a potential $\tilde{V}(\tilde{\phi})$ responsible for the late-time cosmic acceleration. Similar to the case of quintessence DE, in order to explain the DDM evidence, we propose the QDM, which is depicted by a canonical scalar field $\phi$ with a potential $V(\phi)$ that interacts with all the other species only through the standard gravity. The action of QDM reads as
\begin{equation}
S=\int {\rm d}^4x\sqrt{-g}\left[\frac{R-2\Lambda}{2\kappa^2}+\mathcal{L}_{\rm QDM}\right]+S_{M^\prime}, \label{eq:action}
\end{equation}
with
\begin{equation}
\mathcal{L}_{\rm QDM}=-\frac{1}{2}g^{\mu\nu}\partial_\mu\phi\partial_\nu\phi-V(\phi), \label{eq:qdm}
\end{equation}
where $R$, $\Lambda$ and $S_{M^\prime}$ are the Ricci scalar curvature, the cosmological constant, and the action of matter excluding QDM, respectively, and $\kappa^2\equiv8\pi G$, where $G$ is the Newtonian gravitational constant.

\begin{figure*}
	\centering
	\includegraphics[scale=0.6]{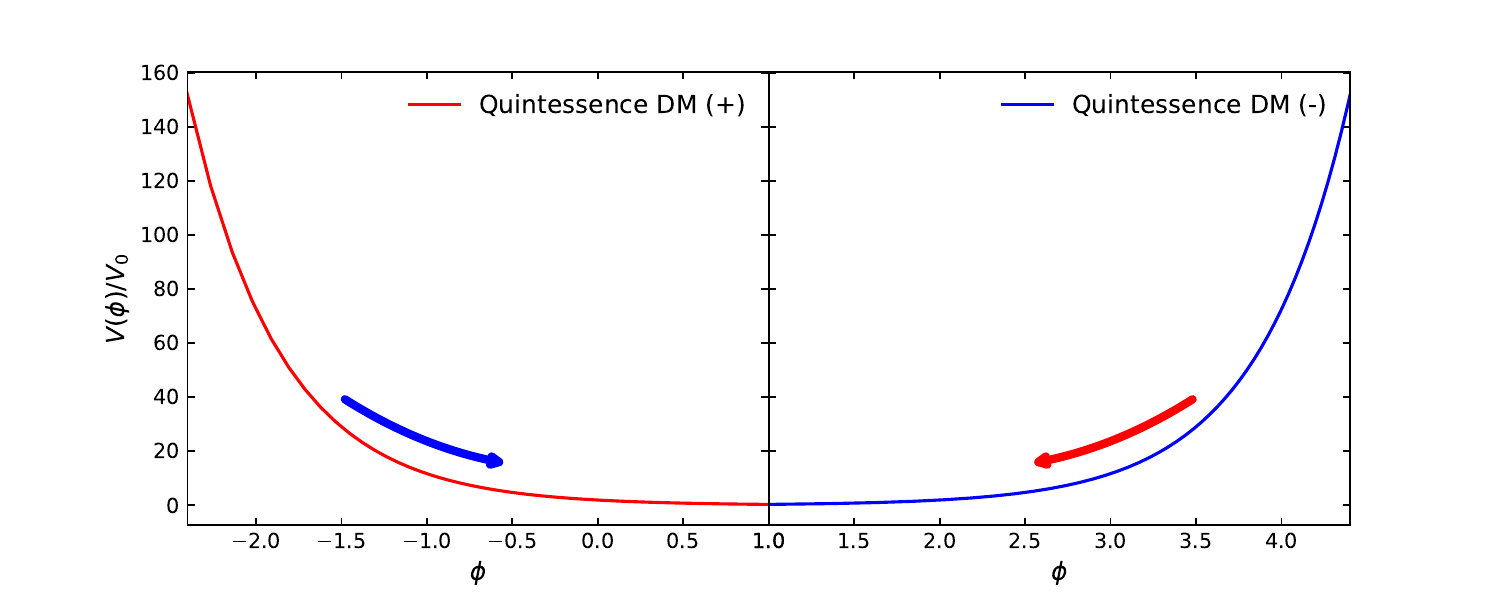}
	\caption{The normalized potential $V(\phi)/V_0$ as a function of $\phi$ in the reconstructed QDM model. The symbols ``$+$'' and ``$-$'' correspond to the plus and minus signs in Eq.(\ref{eq:phi1}), respectively. The arrows indicate the evolution direction of the potential.}\label{fig:vphiphi}
\end{figure*}

\begin{figure*}
	\centering
	\hspace*{-1.7cm}
	\includegraphics[scale=0.52]{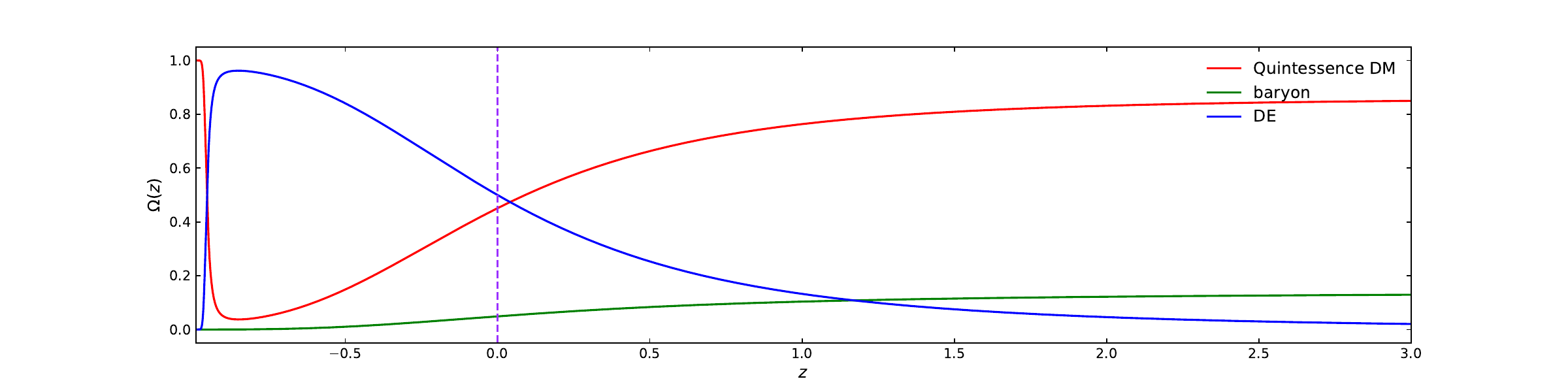}
	\caption{The late-time and future redshift evolution of the fractions of baryons, QDM and DE. The vertical dashed line denotes $z=0$.}\label{fig:Omegaz}
\end{figure*}

In the framework of general relativity \cite{Einstein:1916vd}, considering a spatially flat, homogeneous and isotropic FLRW universe, the energy density and pressure of the QDM are expressed as
\begin{equation}
\rho_{\rm QDM}=\frac{1}{2}\dot{\phi}^2+V(\phi), \label{eq:density}
\end{equation}
\begin{equation}
p_{\rm QDM}=\frac{1}{2}\dot{\phi}^2-V(\phi). \label{eq:pressure}
\end{equation}
The Friedmann equations are easily shown as 
\begin{equation}
H^2=\frac{\kappa^2}{3}\left[\frac{1}{2}\dot{\phi}^2+V(\phi)+\rho_{b}+\rho_{de}\right], \label{eq:friedman1}
\end{equation}
\begin{equation}
\dot{H}=-\frac{\kappa^2}{2}\left[\dot{\phi}^2+\rho_{b}+p_{b}+\rho_{de}+p_{de}\right], \label{eq:friedman2}
\end{equation}
where $H$ denotes the cosmic expansion rate at a scale factor $a$ and $\rho$ and $p$ are the mean energy densities and pressures of different species including baryons, QDM and DE. Here we ignore the radiation and neutrino components, since they hardly contribute to the background evolution of the universe at late times. The variation of the action Eq.(\ref{eq:action}) with respect to $\phi$ gives the so-called Klein-Gordon equation of the QDM 
\begin{equation}
\ddot{\phi}+3H\dot{\phi}+V^{\prime}(\phi)=0, \label{eq:kge}
\end{equation}
where the dot and the prime denote the derivatives with respect to the cosmic time $t$ and the field $\phi$, respectively.

\section*{\normalsize III. Quintessence reconstruction of DDM}
Employing the EoS of the SDDM (single-parameter DDM) model proposed in our previous work \cite{Wang:2025zri}, $\omega_{dm}(a)=\omega_{dm}a$, and its energy density, $\rho_{dm}(a)=\rho_{dm0}a^{-3}{\rm e}^{3\omega_{dm}(1-a)}$, the QDM's energy density and pressure can be written as
\begin{equation}
\frac{1}{2}\dot{\phi}^2+V(\phi)=\rho_{dm0}a^{-3}{\rm e}^{3\omega_{dm}(1-a)}, \label{eq:qdens}
\end{equation}   
\begin{equation}
\frac{1}{2}\dot{\phi}^2-V(\phi)=\omega_{dm}\rho_{dm0}a^{-2}{\rm e}^{3\omega_{dm}(1-a)}, \label{eq:qpres}
\end{equation}   
where $\omega_{dm}$ and $\rho_{dm0}$ are, respectively, today's DM EoS and energy density in the SDDM model (see \cite{Wang:2025zri} for details). After some simple algebraic manipulations, we obtain
\begin{equation}
\dot{\phi}^2=(1+\omega_{dm}a)\rho_{dm0}a^{-3}{\rm e}^{3\omega_{dm}(1-a)}, \label{eq:phi}
\end{equation}  
\begin{equation}
V(\phi)=\frac{1}{2}(1-\omega_{dm}a)\rho_{dm0}a^{-3}{\rm e}^{3\omega_{dm}(1-a)}. \label{eq:vphi}
\end{equation}  
Subsequently, it is easy to see that
\begin{equation}
\frac{{\rm d}\phi}{{\rm d}a}=\pm\left[\frac{3(1+\omega_{dm}a)\Omega_{dm0}a^{-5}{\rm e}^{3\omega_{dm}(1-a)}}{\Omega_{b0}a^{-3}+\Omega_{dm0}a^{-3}{\rm e}^{3\omega_{dm}(1-a)}+1-\Omega_{b0}-\Omega_{dm0}}\right]^{\frac{1}{2}}, \label{eq:phi1}
\end{equation}  
\begin{equation}
\frac{V(\phi)}{V_0}=\frac{1}{2}(1-\omega_{dm}a)\Omega_{dm0}a^{-3}{\rm e}^{3\omega_{dm}(1-a)}. \label{eq:vphi1}
\end{equation}  

To solve these two equations, for simplicity, we use the initial condition $\phi_{a=1}=M_{\rm pl}\equiv1/\sqrt{8\pi G}$, where $M_{\rm pl}$ is the reduced Planck mass. Consequently, Eq.(\ref{eq:friedman1}) is rewritten as $H^2=\rho/(3M_{\rm pl}^2)$. Without loss of generality, it is conveniently to define the present-day potential value as $V_0\equiv\rho_0=3H_0^2M_{\rm pl}^2$. Here $\rho_0$ is actually today's critical density of the universe.

\begin{figure}
	\centering
	\includegraphics[scale=0.5]{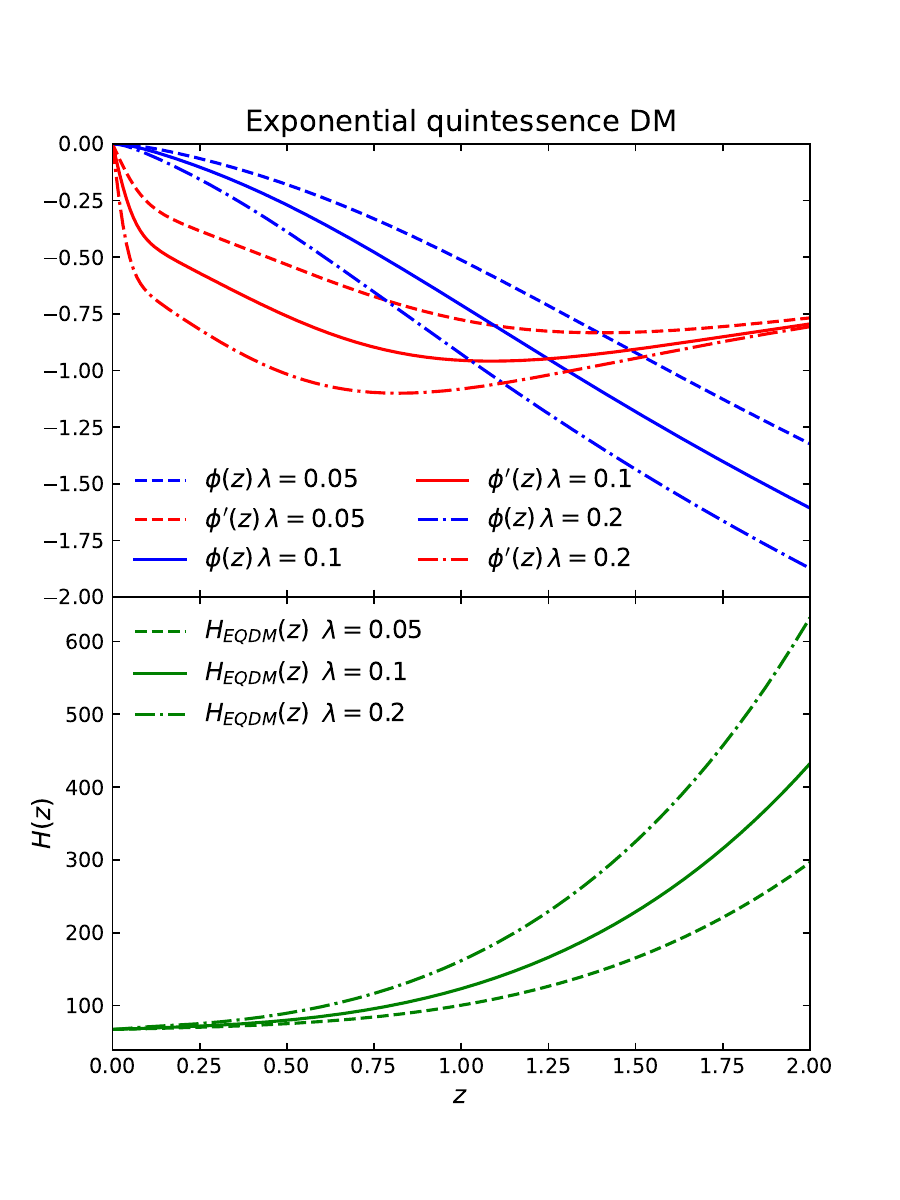}
	\caption{{\it Upper panel.} The field $\phi(z)$ and its derivative $\phi^{\prime}(z)$ as a function of reshift $z$ for different values of $\lambda$ in the EQDM model. {\it Lower panel.} The Hubble parameter $H(z)$ as a function of reshift $z$ for different values of $\lambda$ in the EQDM model.}\label{fig:EQDM}
\end{figure}

Using the consistent values $\omega_{dm}=-0.15$, $\Omega_{dm0}=0.451$, $\Omega_{b0}=0.049$ and $\Omega_{de0}=0.5$ with CMB-only constraints on the SDDM model at $1\,\sigma$ confidence level, we calculate the reconstructed field value $\phi$ and the normalized potential $V(\phi)/V_0$. The results are shown in Fig.\ref{fig:vphiphi}. It is easy to find that the potential of QDM monotonically decreases over time and finally reaches the value $V(\phi_{a=1})/V_0=0.2593$. From Eqs.(\ref{eq:phi1}) and (\ref{eq:vphi1}), one can also see that today's DM EoS and DM fraction in the SDDM model can significantly affect the reconstructed field and potential. Furthermore, in Fig.\ref{fig:Omegaz}, we exhibit the evolution of the fractions of different species including the QDM in the redshift range $z\in(-1,\,3]$. We find that: (i) QDM dominates the evolution of the universe until close to the present day ($z\sim0.05$); (ii) The energy density of DE exceeds that of baryons at $z\sim1.2$; (iii) In the distant future, the DM fraction will experience an extremely fast increase and accordingly the DE fraction will experience an extremely fast decrease; (iv) The ultimate universe is completely dominated by DM, namely $\Omega_{dm}(z=-1)=1$ (see \cite{Wang:2025zri}).

\section*{\normalsize III. Exponential QDM}
Besides the scalar field reconstruction of DDM, we also consider a viable candidate of QDM, i.e., the exponential quintessence DM (EQDM)
\begin{equation}
V(\phi)=V_0{\rm e}^{\kappa\lambda\phi}, \label{eq:eqdm}
\end{equation} 
where $\lambda$ is the sole parameter that characterizes the background dynamics of the EQDM. Specifically, we numerically solve Eq.(\ref{eq:kge}) by assuming $\lambda=0.05$, $0.1$ and $0.2$, respectively. The initial conditions used here are $\phi(z=0)=\phi^{\prime}(z=0)=0$. The corresponding results are shown in Fig.\ref{fig:EQDM}. One can easily find that: (i) With increasing values of $\lambda$, $\phi(z)$ and $\phi^{\prime}(z)$ become smaller at each redshift, while the cosmic expansion rate are larger; (ii) The field value of $\phi$ monotonically increases with the evolution of the universe. Note that here we just take the EQDM as a case study to explore the nature of DDM. Actually, besides the scalar field considered here, there will be many theoretical possibilities that can explain the dynamical evolution of DM, e.g., modified gravities. 

It is noteworthy that the numerical solutions shown in Fig.\ref{fig:EQDM} just give simple behaviors of QDM and are not consistent with current observations. To give compatible solutions with data, one needs to use reasonable initial conditions and viable $\lambda$ in the potential energy. Since we have implemented the quintessence reconstruction of DDM that is well consistent with observations \cite{Wang:2025zri}, a convenient way is directly fitting the curves in Fig.\ref{fig:EQDM} with the exponential potential $V(\phi)=V_0{\rm e}^{\kappa\lambda\phi}$. For the plus and minus signs in Eq.(\ref{eq:phi1}), we find $\lambda=1.059$ and $-2.113$, respectively. These two solutions can well produce the evolution of the matter fraction show in Fig.\ref{fig:Omegaz}.

\section*{\normalsize IV. Discussions and conclusions}
The $\Lambda$CDM model states that the late-time universe mainly consists of two components, i.e., the CDM and the cosmological constant. However, the recent cosmological observations reveal that the nature of the dark sector of the universe should likely be the coexistence of DDM and DDE, which contradicts the traditional understanding of the universe's evolution and composition. 

In this short study, we explain more theoretically the redshift evolution of DDM that is firstly identified with the latest cosmological observations in our previous work \cite{Wang:2025zri}. Specifically, we carry out a scalar field reconstruction to find a quintessence field that is well consistent with observations. Very interestingly, we find the reconstructed solution can be well described by an exponential potential. Therefore, we conclude that the exponential quintessence can act as a viable candidate of DDM. In light of the fact that cosmological observations strongly prefer that DE is quintessential, the nature of the dark sector of today's universe should likely be both dynamical and quintessential.

\section*{\normalsize Acknowledgements}
DW is supported by the CDEIGENT fellowship of Consejo Superior de Investigaciones Científicas (CSIC). DW warmly thanks the hospitality of the Centro de Ciencias de Benasque Pedro Pascual during the preparation of this study.


\begin{thebibliography}{99}
\bibitem{Wang:2025zri}
D.~Wang,
``Evidence for Dynamical Dark Matter,''
[arXiv:2504.21481 [astro-ph.CO]].

\bibitem{SupernovaSearchTeam:1998fmf}
A.~G.~Riess \textit{et al.} [Supernova Search Team],
``Observational evidence from supernovae for an accelerating universe and a cosmological constant,''
Astron. J. \textbf{116}, 1009-1038 (1998).

\bibitem{SupernovaCosmologyProject:1998vns}
S.~Perlmutter \textit{et al.} [Supernova Cosmology Project],
``Measurements of $\Omega$ and $\Lambda$ from 42 high redshift supernovae,''
Astrophys. J. \textbf{517}, 565-586 (1999).

\bibitem{DESI:2024mwx}
A.~G.~Adame \textit{et al.} [DESI],
``DESI 2024 VI: Cosmological Constraints from the Measurements of Baryon Acoustic Oscillations,''
[arXiv:2404.03002 [astro-ph.CO]].

\bibitem{Planck:2018vyg}
N.~Aghanim \textit{et al.} [Planck],
``Planck 2018 results. VI. Cosmological parameters,''
Astron. Astrophys. \textbf{641}, A6 (2020)
[erratum: Astron. Astrophys. \textbf{652}, C4 (2021)].

\bibitem{ACT:2023kun}
M.~S.~Madhavacheril \textit{et al.} [ACT],
``The Atacama Cosmology Telescope: DR6 Gravitational Lensing Map and Cosmological Parameters,''
Astrophys. J. \textbf{962}, no.2, 113 (2024).

\bibitem{Brout:2022vxf}
D.~Brout \textit{et al.},
``The Pantheon+ Analysis: Cosmological Constraints,''
Astrophys. J. \textbf{938}, no.2, 110 (2022).

\bibitem{Rubin:2023ovl}
D.~Rubin \textit{et al.},
``Union Through UNITY: Cosmology with 2,000 SNe Using a Unified Bayesian Framework,''
[arXiv:2311.12098 [astro-ph.CO]].


\bibitem{DES:2024jxu}
T.~M.~C.~Abbott \textit{et al.} [DES],
``The Dark Energy Survey: Cosmology Results with \ensuremath{\sim}1500 New High-redshift Type Ia Supernovae Using the Full 5 yr Data Set,''
Astrophys. J. Lett. \textbf{973}, no.1, L14 (2024).

\bibitem{DESI:2024uvr}
A.~G.~Adame \textit{et al.} [DESI],
``DESI 2024 III: Baryon Acoustic Oscillations from Galaxies and Quasars,''
[arXiv:2404.03000 [astro-ph.CO]].

\bibitem{DESI:2024lzq}
A.~G.~Adame \textit{et al.} [DESI],
``DESI 2024 IV: Baryon Acoustic Oscillations from the Lyman Alpha Forest,''
[arXiv:2404.03001 [astro-ph.CO]].

\bibitem{DESI:2025zgx}
M.~Abdul Karim \textit{et al.} [DESI],
``DESI DR2 Results II: Measurements of Baryon Acoustic Oscillations and Cosmological Constraints,''
[arXiv:2503.14738 [astro-ph.CO]].

\bibitem{DESI:2025fii}
K.~Lodha \textit{et al.} [DESI],
``Extended Dark Energy analysis using DESI DR2 BAO measurements,''
[arXiv:2503.14743 [astro-ph.CO]].

\bibitem{DESI:2025zpo}
M.~Abdul Karim \textit{et al.} [DESI],
``DESI DR2 Results I: Baryon Acoustic Oscillations from the Lyman Alpha Forest,''
[arXiv:2503.14739 [astro-ph.CO]].

\bibitem{Wang:2025bkk}
D.~Wang and D.~Mota,
``Did DESI DR2 truly reveal dynamical dark energy?,''
[arXiv:2504.15222 [astro-ph.CO]].

\bibitem{Chevallier:2000qy}
M.~Chevallier and D.~Polarski,
``Accelerating universes with scaling dark matter,''
Int. J. Mod. Phys. D \textbf{10}, 213-224 (2001).

\bibitem{Linder:2002et}
E.~V.~Linder,
``Exploring the expansion history of the universe,''
Phys. Rev. Lett. \textbf{90}, 091301 (2003).

\bibitem{Wang:2025owe}
D.~Wang,
``Questioning Cosmic Acceleration with DESI: The Big Stall of the Universe,''
[arXiv:2504.15635 [astro-ph.CO]].

\bibitem{Fujii:1982ms}
Y.~Fujii,
``Origin of the Gravitational Constant and Particle Masses in Scale Invariant Scalar - Tensor Theory,''
Phys. Rev. D \textbf{26}, 2580 (1982).

\bibitem{Ford:1987de}
L.~H.~Ford,
``Cosmological Constant Damping by Unstable Scalar Fields,''
Phys. Rev. D \textbf{35}, 2339 (1987).

\bibitem{Wetterich:1987fm}
C.~Wetterich,
``Cosmology and the Fate of Dilatation Symmetry,''
Nucl. Phys. B \textbf{302}, 668-696 (1988).

\bibitem{Ratra:1987rm}
B.~Ratra and P.~J.~E.~Peebles,
``Cosmological Consequences of a Rolling Homogeneous Scalar Field,''
Phys. Rev. D \textbf{37}, 3406 (1988).

\bibitem{Caldwell:1997ii}
R.~R.~Caldwell, R.~Dave and P.~J.~Steinhardt,
``Cosmological imprint of an energy component with general equation of state,''
Phys. Rev. Lett. \textbf{80}, 1582-1585 (1998).


\bibitem{Einstein:1916vd}
A.~Einstein,
``The foundation of the general theory of relativity,''
Annalen Phys. \textbf{49}, no.7, 769-822 (1916).












































\end{thebibliography}
\end{document}